\documentclass[prl,showpacs,twocolumn,aps,superscriptaddress]{revtex4}
\usepackage{color}

\usepackage{amsmath}
\usepackage{bm}
\usepackage{epsfig}
\usepackage{amssymb}

\usepackage{graphicx}
\usepackage{epsfig}
\usepackage{psfrag}

\newcommand{\ee}{\text{e}}

\newcommand{\tobs}{t_\mathrm{obs}}
\newcommand{\sig}{\sigma}
\newcommand{\TK}{T_\mathrm{K}}
\newcommand{\Td}{T_\mathrm{d}}
\newcommand{\To}{T_\mathrm{o}}
\newcommand{\qEA}{q_\mathrm{EA}}

\begin{document}

\title{Metastable states and space-time
phase transitions in a spin-glass model}

\author{Robert L. Jack}
\affiliation{Department  of Physics, University of Bath, Bath, BA2 7AY, 
United Kingdom}

\author{Juan P. Garrahan}
\affiliation{School of Physics and Astronomy, University of
Nottingham, Nottingham, NG7 2RD, UK}

\begin{abstract}
We study large deviations of the dynamical activity in the random orthogonal model (ROM).  
This is a fully connected spin-glass model with one-step replica symmetry breaking behaviour, 
consistent with the random first-order transition scenario for structural glasses.   We show that this model 
displays dynamical (space-time) phase-transitions between active and inactive phases, 
as demonstrated by singularities in large deviation functions.  
We argue that such transitions are generic in systems with long-lived metastable states.
\end{abstract}


\maketitle

\noindent
Glass transitions and glassy dynamics occur in a wide range
of systems, including structural glasses~\cite{glass}, colloidal 
suspensions,
granular media~\cite{colloids-granular} and spin glasses~\cite{spin-glass}.  
As their glass transitions are approached, the relaxation
in these systems slows down dramatically but their structure remains disordered.  
The increasing relaxation time is often assumed to be a consequence of an 
underlying (continuous) phase transition~\cite{MCT,KTW,Tarjus}, but the existence
of such a transition in structural glasses remains unproven.

We and others have recently proposed that, even if no
thermodynamic phase transition exists in glass formers, 
the underlying transition might be a (discontinuous)
``space-time'' phase transition 
~\cite{Merolle,Garrahan-Fred,Hedges-science}, occurring in trajectory space.
By applying a thermodynamic (large deviation) formalism to ensembles of 
trajectories \cite{Ruelle-book,Touchette}, 
one constructs dynamical free-energies, 
whose singularities can be interpreted as dynamical phase transitions.  
The existence of such first-order transitions can be proven in idealised
lattice models, 
known as kinetically constrained models (KCMs)~\cite{Garrahan-Fred}.
Furthermore, computer simulations reveal behaviour consistent
with these phase transitions in atomistic model 
glass-formers~\cite{Hedges-science}.  
Physically, the idea~\cite{Merolle} is that
the characteristic features of glassy systems arise from
coexistence between active and inactive dynamical phases. 

Here, we consider the
random orthogonal model (ROM)~\cite{ROM-general,
ROM-scaling}, a fully-connected spin-glass model that realises
the one-step replica symmetry breaking (1-RSB)
scenario.  This scenario is the basis for a mean-field theory of structural 
glasses, the random first-order transition theory \cite{KTW}.
We show that the ROM supports coexisting dynamical phases, separated by
first-order space-time phase transitions, as in KCMs.
We argue that these transitions 
occur in systems with long-lived metastable states, including 
generic 1-RSB models.

\begin{figure}
\includegraphics[width=7cm]{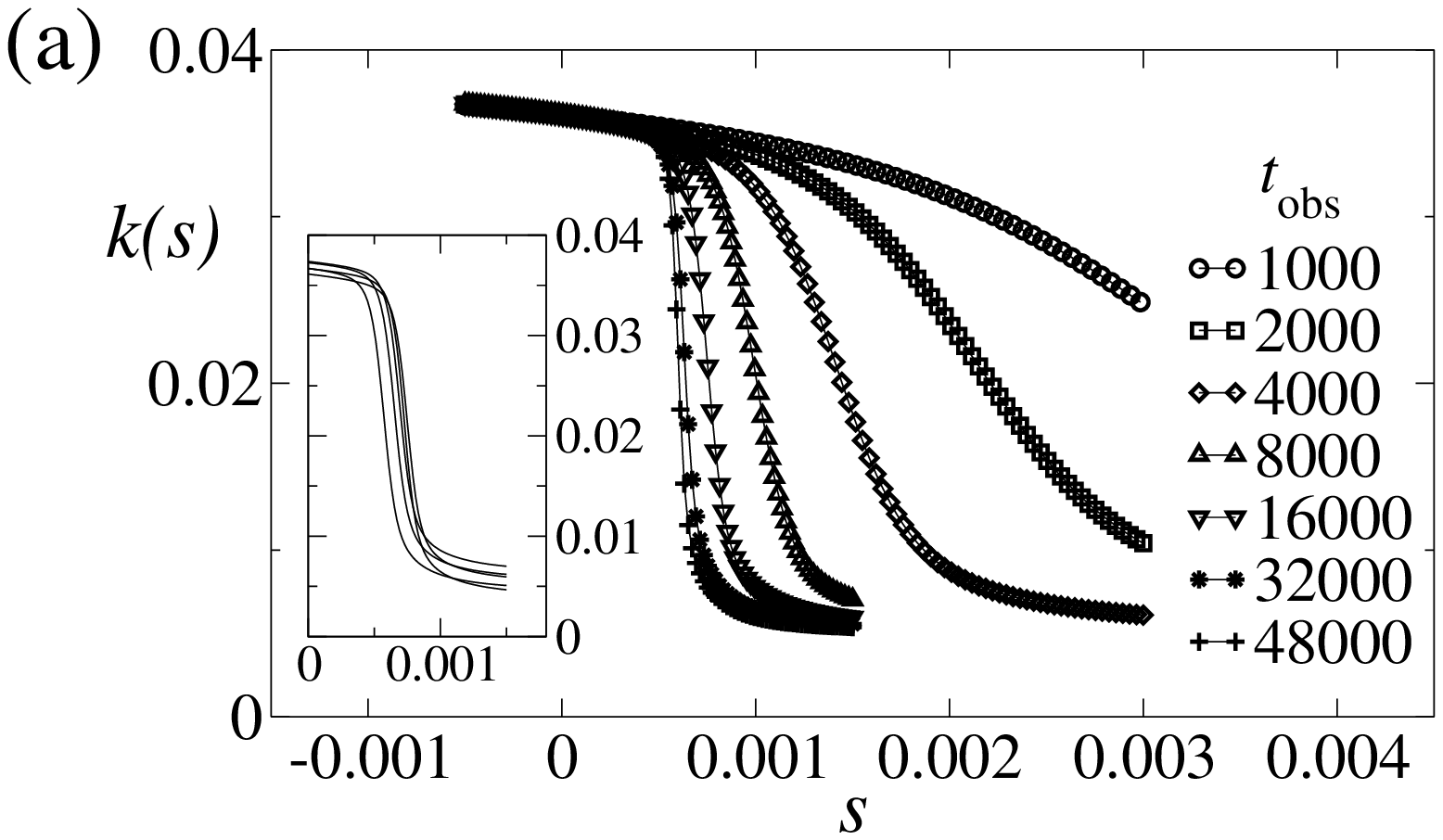}
\includegraphics[width=7cm]{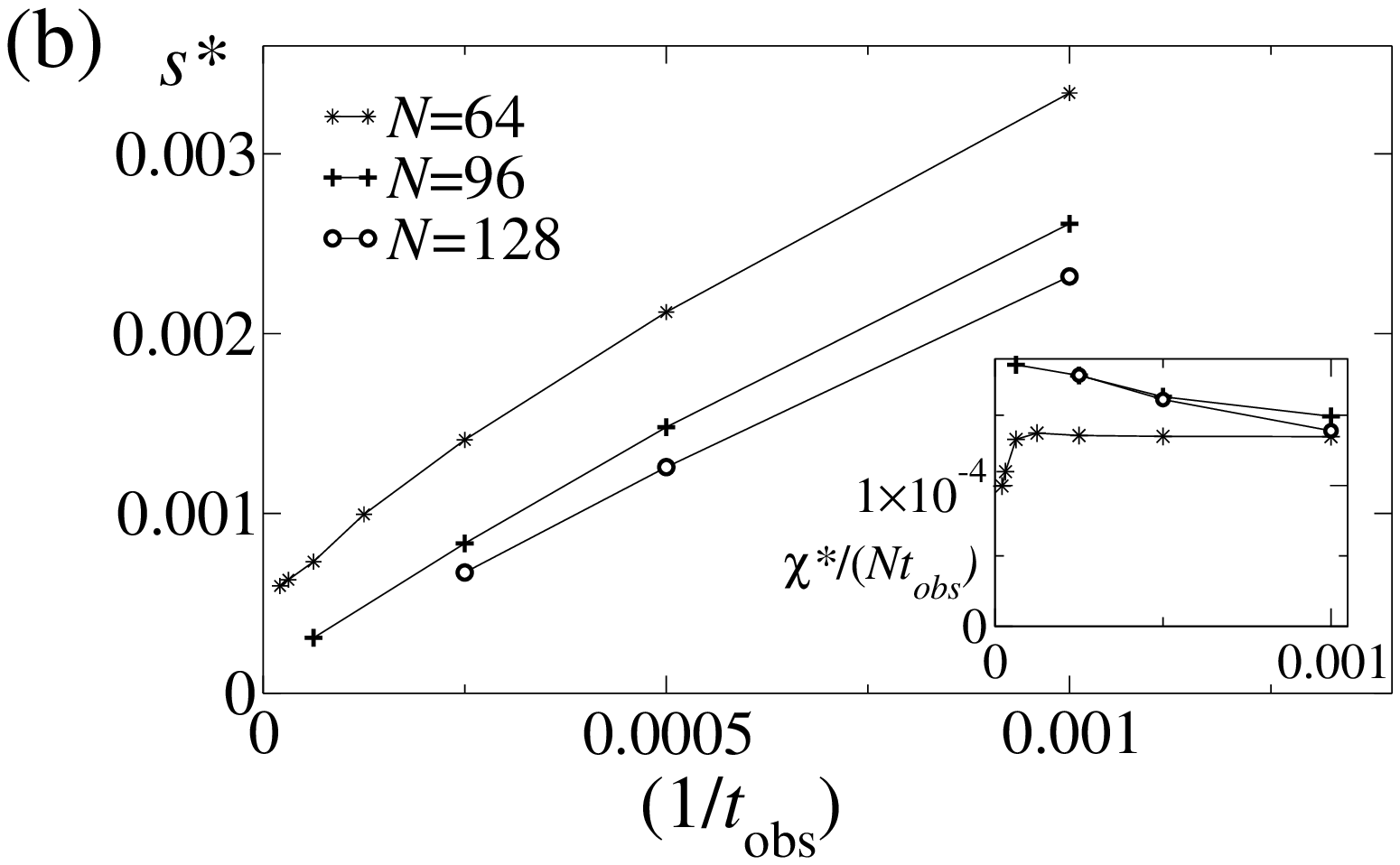}
\caption{{\bf Space-time phase transition in the ROM}.   
(a)~Average activity $k(s)$ as a function of $s$ for $N=64$ and fixed disorder at $T=1/5 > \Td$.  The equilibrium relaxation time 
at this temperature is $\tau\approx110$ (in units of MC sweeps).  
The crossover in $k(s)$ becomes increasingly sharp as $\tobs$ increases.  
The inset to (a) shows $k(s)$ for five different disorder
realisations for $\tobs=16000$.  (b)~Dependence 
of $s^*$ and $\chi^*$ (inset) on observation time and system size.  
This scaling compatible with a space-time phase transition at $s=0$.
}
\label{fig1}
\end{figure}

We apply thermodynamic methods to
measures of dynamical activity, as described in~\cite{Garrahan-Fred}.
Consider a system of $N$ spins (or $N$ particles),
evolving with stochastic dynamics, at temperature $T$.
We define
\begin{equation}
Z(s,\tobs) = \langle \ee^{-sK} \rangle_{\tobs} ,
\label{Zs}
\end{equation}
where $K$ is 
a measure of activity and the average is taken over trajectories 
that run from an initial time $t=0$ to a final time $t=\tobs$, in
an equilibrated system.  In the ROM, the configuration
space is discrete and we take $K$ to be
the number of changes of configuration (kinks) in the 
trajectory~\cite{Fred-Cecile,Merolle}.  
For large $\tobs$, then $Z(s,\tobs) \sim \ee^{\tobs \psi(s)}$.
The function $\psi(s)$ is a large deviation function, and can be thought of as a ``space-time''
free energy.  
Its singularities are space-time phase transitions: i.e., 
qualitative changes in ensembles of trajectories.   
Interpreting $Z(s,\tobs)$ as the partition function for a biased ensemble
of trajectories (the `$s$-ensemble'), we define expectation values
within this ensemble as $\langle A \rangle_s=Z(s,\tobs)^{-1} \langle 
A \ee^{-sK} \rangle_{\tobs}$.  In particular,
\begin{equation}
k(s)\equiv \frac{1}{N\tobs}\langle K\rangle_s ,
\end{equation}
is the mean activity
in the $s$-ensemble. In the limit of large $\tobs$ then
$k(s)\to-\frac1N \psi'(s)$.  (The ensemble with
$s=0$ is simply the equilibrium ensemble of trajectories.)

\begin{figure}[t]
\includegraphics[width=7cm]{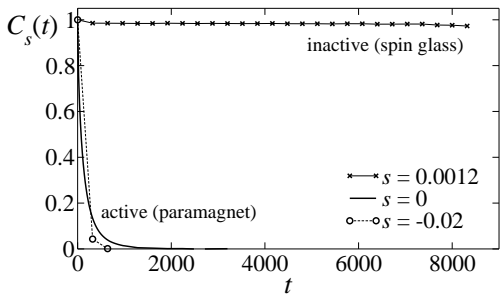}
\caption{{\bf Dynamics in active and inactive phases}.  
Autocorrelation functions $C_s(t)$ in the ROM at $T=1/5$,
for $N=64$ and $\tobs=16000$, illustrating active and inactive phases
obtained by varying $s$.}
\label{fig2}
\end{figure}

The random orthogonal model (ROM) \cite{ROM-general}
consists of $N$ Ising spins $\sig_i=\pm 1$ ($i=1,\ldots,N$) and an energy function
$E=1/2\sum_{i\neq j} J_{ij} \sig_i \sig_j$.   The matrix of quenched random couplings $J_{ij}$
is symmetric and orthogonal.  We construct it as $J = R^T D R$
where $D=\mathrm{diag}(1,-1,1,-1,\dots)$ and $R$ is a randomly
generated $O(N)$ rotation.  Consistent with the 1-RSB scenario in the $N \to \infty$ limit, there are three important
temperatures for the ROM~\cite{ROM-general}:
the static transition temperature
$\TK=0.065$ below which replica symmetry is broken;
the dynamical transition temperature $\Td=0.134$
below which the equilibrium correlation function
has a non-zero limit as $t\to\infty$; and the onset temperature
$\To=0.32$ below which long-lived Thouless-Anderson-Palmer (TAP) states exist~\cite{TAP,
Barrat-pspin,BK}.  

The ROM is straightforwardly simulated using Monte Carlo dynamics.   
Time is measured in Monte Carlo
sweeps throughout, and the only parameter of the model is the temperature $T$.  
We focus first on the regime $\Td < T < \To$ which is the most relevant 
one for supercooled liquids.  
We use transition path sampling~\cite{TPS} to sample
the $s$-ensemble, as described in~\cite{Hedges-science}.
We show results for $N\geq64$ and for representative realisations of the disorder $J_{ij}$.  
Our results depend weakly on the realisation of the disorder, but 
we have not analysed sample-to-sample fluctuations in detail
due the computational effort associated with sampling the $s$-ensemble
(see~\cite{ROM-scaling} for an analysis at equilibrium).

Fig.~\ref{fig1}(a) shows the mean activity $k(s)$ 
in the $s$-ensemble at temperature $T=1/5$.  
Clearly, $k(s)$ 
decreases sharply as $s$ is increased from zero.  
That is, there is a crossover from active behaviour for $s\leq 0$ to inactive
behaviour for larger $s$, and this crossover becomes increasingly sharp 
as $N$ and $\tobs$ are increased.  
The inset to Fig.\ \ref{fig1}(a) suggests that this crossover
is independent of the precise realisation of the disorder $J_{ij}$.  
The susceptibility $\chi(s) \equiv k'(s)$ peaks at the inflection point of the 
curves in panel (a).   
Let $s^*$ be the value of $s$ that maximises $\chi$, and let $\chi^* \equiv \chi(s^*)$ be
the maximal susceptibility.  Fig.\ \ref{fig1}(b) 
shows that $s^*$ decreases towards zero with increasing $N$ and $\tobs$, 
and $\chi^*$ diverges linearly with the space-time volume $N \times \tobs$.  
This finite size scaling is consistent with a sharp (first-order) transition at $s^*=0$.  
We interpret $s=0$ as a line of `dynamical phase coexistence'~\cite{Garrahan-Fred}.

We note that the $s$-ensemble is time-translational invariant (TTI) 
only for times $0\ll t\ll \tobs$, with deviations from TTI
behaviour~\cite{Garrahan-Fred} near the initial and final times.
These boundary effects enhance the contribution of the active
phase to $Z(s,\tobs)$, so that for fixed $N$ we 
expect~\cite{long-paper} that $s^* = s_N + O(1/\tobs)$, consistent with 
Fig.~\ref{fig1}(b).  
The scaling of $\chi^*$ and $s^*$ with $N$ can be accounted for
by considering the lifetimes
of metastable (TAP) states in finite systems.  Briefly,
if $N$ is finite then all metastable states have finite
lifetimes and $\psi(s)$ is analytic for all $s$~\cite{cts-foot}.

We characterise the dynamical behaviour of the ROM in the $s$-ensemble 
via the autocorrelation function, 
$C_s(t) = \langle N^{-1} \sum_i \sig_i(t'+t) \sig_i(t') \rangle_s$, 
which is independent of $t'$ for $0\ll t' < t+t' \ll \tobs$~\cite{Garrahan-Fred}.   
Fig.\ \ref{fig2} shows this function for 
values of $s$ on both sides of the dynamical transition.  
We have $T>T_d$, so the equilibrium dynamics of the ROM are
ergodic, and $C_{s=0}(t)$ decays to zero with a finite relaxation time $\tau$.
For $s<0$, states with high activity dominate the $s$-ensemble
and the trajectories resemble those at equilibrium.
However, for $s>0$, states with low activity predominate, and
Fig.~\ref{fig2} shows that $C_s(t)$ remains finite on the longest time scales
that we can sample.  We define $\qEA \equiv \lim_{t\to\infty} C_s(t)$, with
the limit taken after the limits of large $N$ and $\tobs$. 
We now show that systems realising the 1-RSB scenario
have a first-order dynamical transition from a `paramagnetic state' 
with $\qEA=0$ to a `spin glass' with finite $\qEA$, as $s$ is increased through zero.  
This is consistent with Figs.~\ref{fig1} and~\ref{fig2}, since the ROM
realises this scenario.

Our discussion rests on the existence of a 
large number of metastable states, which can be studied within the TAP approach~\cite{TAP,
Barrat-pspin,BK}.  
The presence of TAP states is sufficient to prove the existence of a space-time phase transition.  
Let $\mathbb W$ be
the master operator associated with the stochastic dynamics of the system,
as in~\cite{Garrahan-Fred}.
Consistent with the 1-RSB scenario,
we assume a separation of time scales, corresponding
to conditions on the eigenspectrum of $\mathbb W$:
There is a spectrum of fast rates larger than some cutoff
$\gamma_\mathrm{f}$ and a spectrum of slow rates smaller than a second
cutoff $\gamma_\mathrm{s} \ll \gamma_\mathrm{f}$.  
On starting in a given configuration, the system relaxes quickly
into a metastable (TAP) state in a time of order $\gamma_\mathrm{f}^{-1}$. 
However, transitions between these states occur much more slowly, 
taking a time of order $\gamma_\mathrm{s}^{-1}$.  
Then, for
$\gamma_\mathrm{f}^{-1} 
\ll  \tobs \ll \gamma_\mathrm{s}^{-1}$, 
the time evolution operator of the system 
is a projection operator onto the TAP states:
\begin{equation}
\ee^{\mathbb W \tobs} = \sum_\alpha | P_\alpha \rangle
\langle Q_\alpha | + O( \ee^{-\gamma_\mathrm{f}\tobs} ) + O( \gamma_\mathrm{s} \tobs) ,
\label{equ:project}
\end{equation}
where $|P_\alpha\rangle$ describes the
(metastable) equilibrium distribution within state $\alpha$, and
 $\langle Q_\alpha|$ gives the probabilities 
of relaxation into state $\alpha$~\cite{Gaveau-Schulman}.  
This result was used in~\cite{BK}, where the trace of 
$\ee^{\mathbb{W}t^*}$
was used to estimate the number of metastable states with lifetimes
greater than $t^*$.  

Now, the partition sum $Z(s,\tobs)$ has a transfer matrix representation 
and the free energy $\psi(s)$ is the largest eigenvalue of
a transfer operator $\mathbb W(s)$~\cite{Lebowitz-Spohn,Fred-Cecile}, 
such that $\mathbb W(0)=\mathbb W$.
Further, the largest eigenvalue of $\mathbb W(s)$ can be estimated 
variationally \cite{Garrahan-Fred}, so that
$\psi(s) \geq \frac{ \langle \Psi | \ee^{\hat{E}/T} 
\mathbb W(s) | \Psi \rangle }{ \langle \Psi | \ee^{\hat{E}/T}
 | \Psi \rangle}$
for any trial state $|\Psi\rangle$,
with $\hat E$ being the energy operator of the system (we take
$k_\mathrm{B}=1$).
For small $s$, we use the variational basis $|P_\alpha\rangle$,
arriving at
\begin{equation}
\psi(s) \geq \psi_\mathrm{var}(s) = -N \min_\alpha [ s k_\alpha ] 
+ O(\gamma_\mathrm{s}) + O(s^2)
\label{equ:var}
\end{equation}
where $k_\alpha$ is the average value of the activity density
$K/(N\tobs)$ for trajectories at
(metastable) equilibrium in state $\alpha$.  
For $\gamma_\mathrm{s} \ll |s| \ll \gamma_\mathrm{f}$, the bound
is saturated~\cite{long-paper}, and
\begin{equation}
k(s) \approx 
\theta(s) \min_\alpha [k_\alpha]   +
\theta(-s) \max_\alpha [k_\alpha ] ,
\label{equ:K-jump}
\end{equation}
where $\theta(s)$ is the step function.  
In the 1-RSB scenario, the slow rate $\gamma_\mathrm{s}$ vanishes
in the limit of  large-$N$.  Taking this limit, followed by a limit
of large $\tobs$, Eq.~(\ref{equ:K-jump}) holds as $|s|\to0$.
Hence, if the states $\alpha$ cover
a finite range of $k_\alpha$ then $k(s)$ is discontinuous at $s=0$. 
Thus, if $\gamma_\mathrm{f}$ is finite and $\gamma_\mathrm{s}\to0$,
there is a first-order dynamical transition at $s=0$,
similar to that seen in KCMs~\cite{Garrahan-Fred}.
The prediction of Eq.\ (\ref{equ:K-jump}) and the numerical observations
of Figs.~\ref{fig1} and \ref{fig2} constitute the key results of this paper:  
for $\Td < T < \To$, the ROM  has a first-order space-time phase transition at $s=0$.  

In thermodynamics,
first-order phase transitions are characterised
by singular responses to boundary fields.  
We take $s=0$ and $\Td<T<\To$, and consider an ensemble 
of trajectories with initial conditions that are equilibrated at 
temperature $T'$.
Within the 1-RSB scenario, the system relaxes
into the equilibrium (active) state for $T'>\Td$, but for
$T'<\Td$ it relaxes into a metastable state with finite 
$\qEA$~\cite{Barrat-pspin}.  In the language
of the $s$-ensemble, the temperature $T'$ corresponds to
a boundary field on the trajectories, and the singular response
at $T'=\Td$ may be linked with wetting phenomena~\cite{wet,
Hedges-science}.

\begin{figure}
\includegraphics[width=7cm]{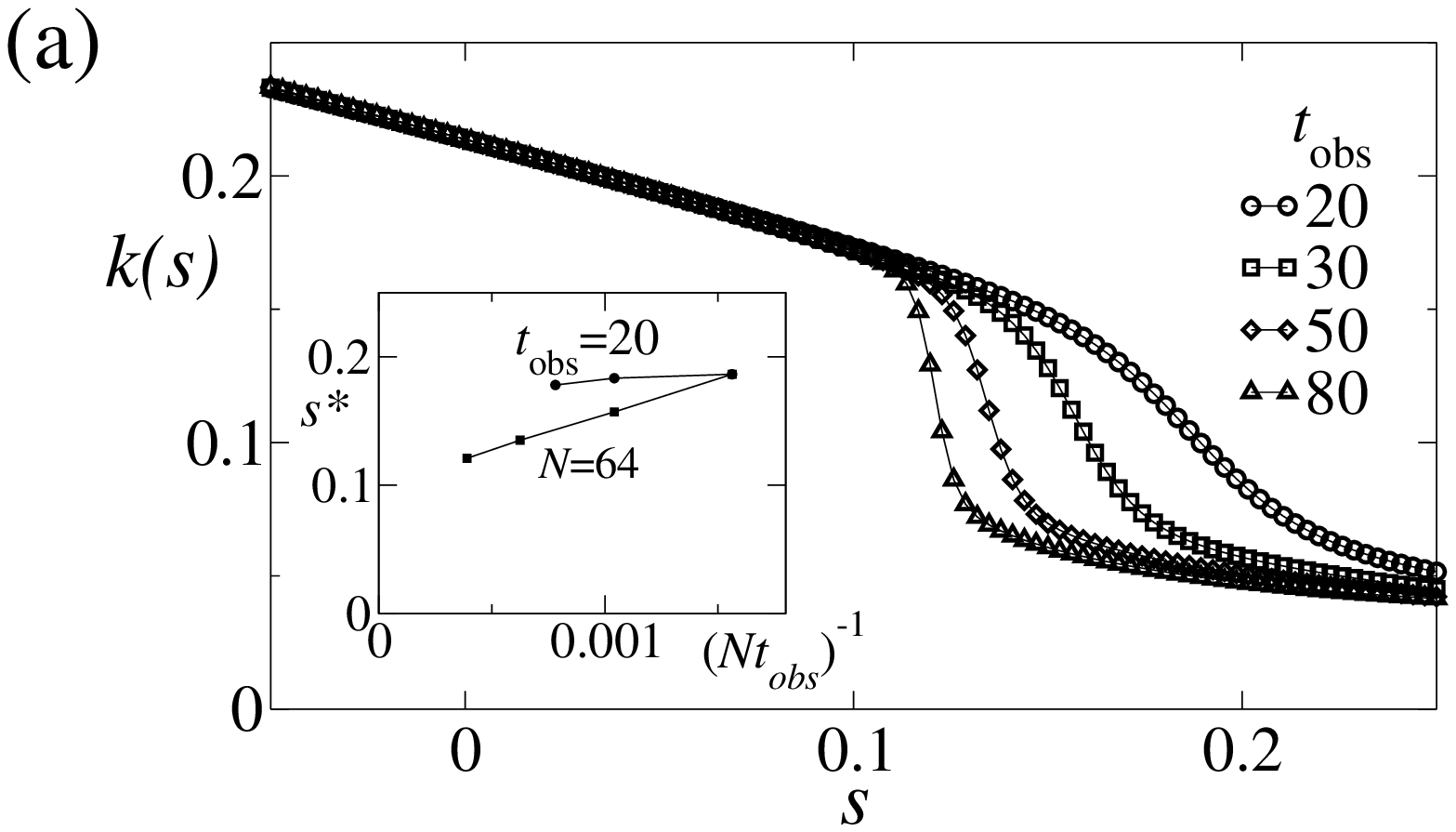}
\includegraphics[width=7cm]{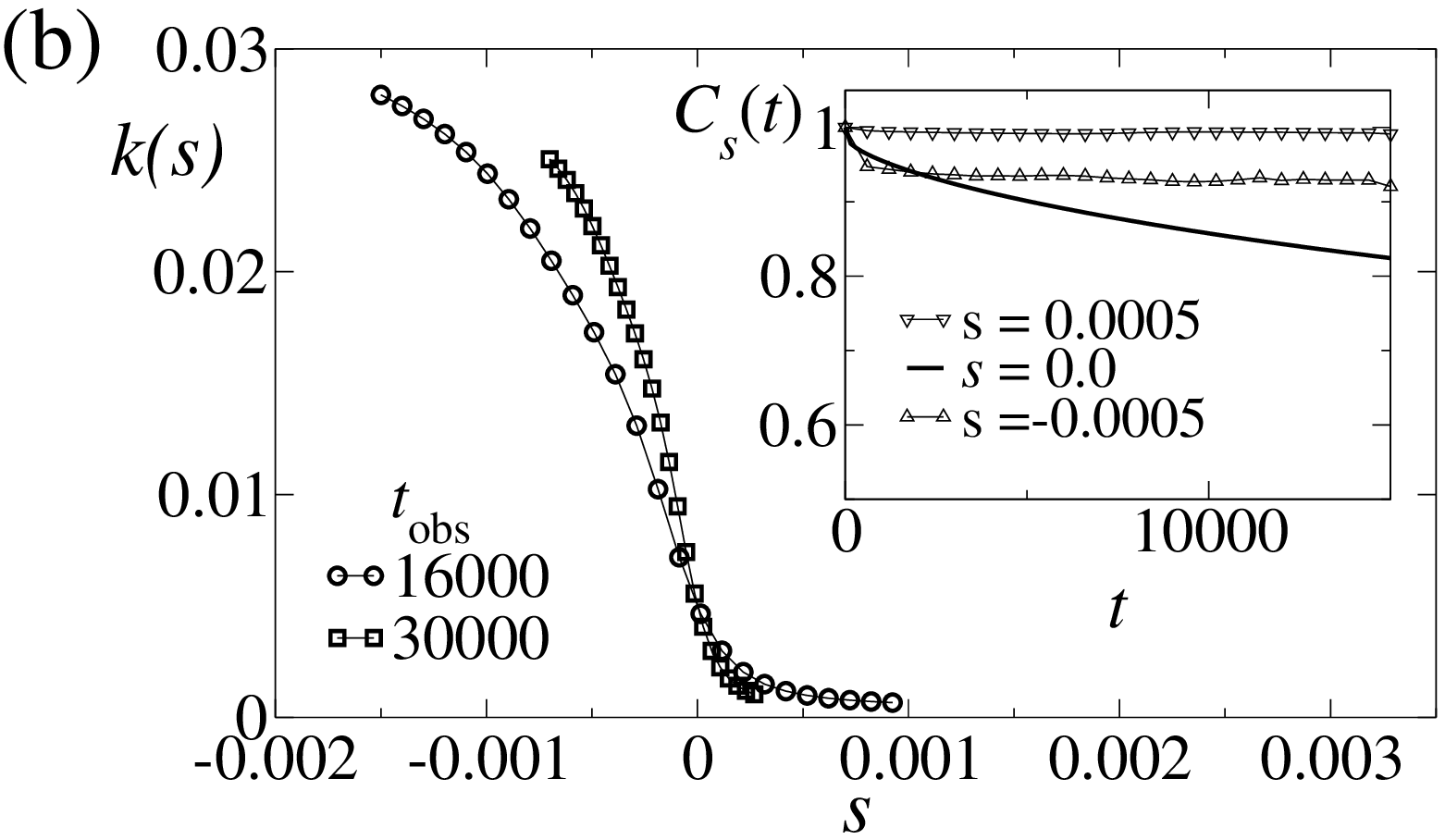}
\caption{{\bf Transitions in the ROM for $T > \To$ and $T < \Td$. }
(a)~Mean activity $k(s)$ as a function of $s$ at $T=1/2>\To$ for increasing $\tobs$ at $N=64$ and fixed disorder, cf.\ Fig.~\ref{fig1}.  The equilibrium correlation time at this temperature is $\tau \approx 4$. Inset: effect on $s^*$ of increasing $N$ and $\tobs$.  These observations are
consistent with a transition at finite $s^*$ in the thermodynamic limit. 
[On increasing $\tobs$ at $N=64$, $\chi^*/(N\tobs)$ increases weakly (not shown).]
(b)~Mean activity $k(s)$ at $T=1/9<\Td$ for $N=64$.  
The behaviour of $k(s)$ is consistent with a first-order transition at $s^*=0$.  
Inset: autocorrelation $C_s(t)$ for $\tobs=3\times10^4$ for various $s$.  
The relaxation time at $s=0$ is $\tau\approx10^7$, although this depends strongly on system size
since $T<\Td$.}
\label{fig3}
\label{fig4}
\end{figure}

So far we have considered only $\Td<T<\To$.
For temperatures above the onset temperature, $T>\To$, metastable states are no longer infinitely 
long-lived and the slow rate $\gamma_\mathrm{s}$ remains finite even as $N\to\infty$.
It follows that $k(s)$ is continuous at $s=0$.
In the absence of a diverging slow
time scale associated with the operator $\mathbb W$, 
one might expect $k(s)$ to be analytic for all $s$~\cite{cts-foot}.
However, for $T>\To$,
analytic arguments~\cite{long-paper} and numerical results both 
indicate a first-order dynamic phase transition between 
active and inactive phases that occurs at finite $s^*$. 
Fig.~\ref{fig3}(a) shows the numerical evidence for this transition.  
Dynamical phase transitions at finite $s$ have been found in other spin
models for which all states have finite lifetimes~\cite{Fred-Cecile,Rob-Fred}.

\begin{figure}
\includegraphics[width=5.0cm]{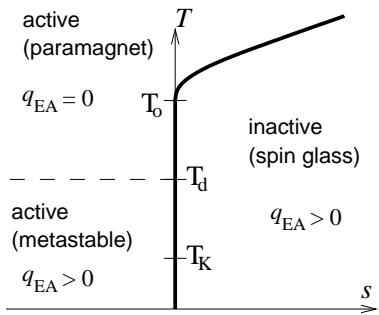}
\caption{{\bf Proposed space-time phase diagram}. 
The heavy line is a first-order transition between active and inactive dynamical phases.  
We expect dynamical phase coexistence at $s=0$ in 1-RSB systems 
for all temperatures below the onset temperature $\To$.  
For $T>\To$ coexistence takes place at $s>0$.  
The dashed line separates the metastable active
state of Fig.~\ref{fig4}(b) from the paramagnetic active state of Fig.~\ref{fig1}.
 }
\label{fig5}
\end{figure}

For $T < \Td$, the behaviour is 
subtle and we postpone a detailed discussion to later work~\cite{long-paper}.
In this regime, 1-RSB systems have `threshold' states which are associated with
aging behaviour~\cite{CuKu}.  
The relaxation time within the paramagnetic state diverges,
and the `gap' $(\gamma_\mathrm{f}-\gamma_\mathrm{s})$ vanishes.
We evaluate the bound in Eq.~(\ref{equ:var}) while excluding the
paramagnetic state from the minimisation.  The resulting bounds may
not be saturated so the proof of Eq.~(\ref{equ:K-jump}) fails.
However, as long as the minimisation contains states with
a finite range of $k_\alpha$, Eq.~(\ref{equ:var}) establishes the existence of
a first-order space-time phase transition for $T<\Td$,
similar to that for $T>\Td$.
Fig.~\ref{fig4}(b) shows numerical results consistent with such a transition.
Note that $\qEA$ remains finite for $s<0$, suggesting that the active state is constructed from active metastable states and not from paramagnetic `threshold' states.

The dynamical phase structure of the ROM is summarised in the $(s,T)$ phase diagram of Fig.~\ref{fig5}.
For temperatures between the dynamical transition temperature and the onset of metastability, 
$\Td<T<\To$,  metastable states lead to a first-order dynamical
phase transition at $s=0$ (Fig.~\ref{fig1}).  Thus, the equilibrium ensemble of
trajectories is associated with coexistence between active (ergodic) and inactive phases.
Above $\To$, all metastable states in the model have finite lifetimes,
and the coexistence line moves to finite $s$, Fig.~\ref{fig3}(a).  
For $T < \Td$,  the first-order transition remains at $s=0$ but it now separates dynamics 
within metastable states with high and
low activity, Fig.~\ref{fig4}(b).  This suggests that for $s<0$ there is a transition near $\Td$ 
between an active ergodic phase with $\qEA=0$ and an active 
but non-ergodic phase in which the activity $k(s)$ is larger than its equilibrium
value $k(0)$ but $\qEA>0$~\cite{Rob-Fred}.  
At $T_K$ the system undergoes an `entropy crisis': for $T<\TK$, the
TAP states are numerous although their associated entropy (complexity) vanishes.
Nevertheless, our arguments for $T<\Td$ still apply,
indicating that the transition remains at $s=0$. 

We have focussed on the ROM in this article, but
Eqs.~(\ref{equ:var}) and (\ref{equ:K-jump}) indicate that phase diagrams for generic
1-RSB systems should be similar to Fig.~\ref{fig5}. How this
picture differs between mean-field and finite-dimensional
systems is an important open question.
Our main conclusion is that
dynamical phase coexistence between active and inactive phases
is not restricted to idealised KCMs [8] but also present in atomistic liquids [9], and,
as we have shown here, in spin-glasses.
The s-ensemble is the natural method for studying inactive and metastable
states and their consequences in glassy systems in general.

We thank D. Chandler and F. van Wijland for helpful discussions, 
and RLJ thanks D.R. Reichman for suggesting the suitability of the ROM 
for numerical simulations.  During the early stages of this work, RLJ was funded
through the U.S. Office of Naval Research through grant 
No.~N00014-07-1-068.


\begin{thebibliography}{99}
\bibitem{cts-foot}
In general,
$k(s)$ and its derivatives are smooth if 
we take $\tobs\to\infty$ at finite $N$. 
Thus, the crossover in Fig.~\ref{fig1}(a) occurs at a value of $s^*$
that extrapolates to $s_N>0$ in this limit, while
$\chi^*/(N\tobs)\to0$ for large enough $\tobs$.  This latter quantity
has a finite limit only if both $N$ and $\tobs$
are taken to infinity together.
Indeed, Fig.~\ref{fig1}(b) shows that $\chi^*/(N\tobs)$
decreases for very large $\tobs$ and $N=64$.
\end{thebibliography}
\end{document}